\pdfoutput=1

\documentclass[aps,preprintnumbers,amsmath,amssymb,twocolumn, tightenlines,superscriptaddress,nofootinbib]{revtex4}
\usepackage{graphicx}
\usepackage{slashed}
\usepackage{lipsum}
\usepackage{tikz,mathpazo}
\usetikzlibrary{shapes.geometric, arrows}
\usepackage[none]{hyphenat}
\usepackage[english]{babel}

\begin{document}


\title{Proca equation and vector field quantization in rotating system}

\author {Tian Xu} 
\address{Physics Department, Beihang University, 37 Xueyuan Rd, Beijing 100191, China}

\author {Yin Jiang} 
\address{Physics Department, Beihang University, 37 Xueyuan Rd, Beijing 100191, China}
\date{\today}

\begin{abstract}
A strong background field will change the vacuum structure and the proper basis of a system drastically in both classical and quantum mechanics, e.g. the Landau levels in a background magnetic field. The situation is the same for the rotating case. In such a system the usual set of plane-wave states would no longer be suitable as a starting point of perturbation. Alternatively and straightforwardly in a rapidly and globally rotating system, it is better  to reformulate the perturbation computation in principle.  In this work we will complete the first step for the spin-1 field, which includes solving the Proca equation in present of a background rotation and complete its canonical quantization. It will be shown that because of the symmetry the eigen states are actually the same as the ones of Maxwell equations in cylindrical coordinate. The propagator as well as the near-central approximation will be obtained by considering the vorticity areas are so small in the relativistic QGP. 
\end{abstract}

\pacs{12.38.Aw, 12.38.Mh}
\maketitle

\section{Introduction}
As the measurements at the Relativistic Heavy Ion Collider(RHIC) and the Large Hadron Collider(LHC) processing\cite{BedangadasMohantyfortheALICE:2017xgh,Zhou:2019lun,Mohanty:2020bqq}, the polarization behaviors of vector mesons are more and more non-trivial and difficult to understand recently. These measurements which were motivated by explorations of large background field in the fire ball of Quark Gluon Plasma(QGP) produced in the relativistic heavy ion collisions, includes the magnetic and vorticity fields, have led the studies of the QGP phenomenological properties into a more complicated and mysterious situation. For the magnetic field, although the mechanism of the chiral magnetic effect(CME) is quite clear theoretically\cite{Fukushima:2008xe,Kharzeev:2013ffa,Sadofyev:2010pr}, its detection appears to be quite difficult because of large non-relevant fluctuations\cite{Wang:2012qs,Aziz:2020nia,Haque:2018jht,Li:2020dwr}. While for the vorticity, after a inspiring qualitative agreements of the simulations of the $\Lambda$ polarization dependence on the collision energy and centrality with the STAR measurements\cite{STAR:2017ckg,Xie:2017upb}, it indicated that the vorticity distribution and evolution in QGP may be much more out of ones' expectation by both the quantitative analysis on the global polarization and the qualitative mismatch of the local polarization profiles\cite{Shapoval:2017jej,Li:2017dan, Becattini:2020ngo}.

The dependence of the vorticity amplitude on the collision energy and centrality has been studied in \cite{Wei:2018zfb} for the initial state of QGP and further simulated in \cite{Jiang:2016woz} with a multi-phase transport model(AMPT) for the whole period of QGP evolution. As the qualitative predictions is quite reasonable, ones believe that the traditional understanding  on the vorticity should not be completely wrong.  Thus from the aspects of kinetic theory, quantum Wigner functions and hydrodynamics, various attempts\cite{Kharzeev:2013ffa,Li:2017dan,Karpenko:2021wdm,Gao:2019znl,Weickgenannt:2019dks,DelZanna:2013eua,Karpenko:2013wva} have been made to mend the usual theory which is basing on the straightforward scenario of the parton collisions, such as different vorticity definitions by considering the thermal environment\cite{Becattini:2020ngo,Becattini:2015ska} and novel vacuum structures by the strongly rotating system\cite{Fang:2016uds}. In most of these works, quraks, which serve as the visible spin-carries of the final-state hadrons, have attracted almost all the interests\cite{Yang:2017sdk,Sheng:2019kmk}. The gluons, which carrying double spins and thus suffered double polarization effects, are neglected because of the technical problems in most cases.  As it is closely related with the quantum chromodynamics(QCD)'s fundamental problem, in this work we will only take our first step towards the gluon's part, namely to study the vector field in the present of a background rotation field. 

A strong background field will change the vacuum structure of a system drastically in both classical and quantum mechanics\cite{Fukushima:2018grm}. The Landau level is the most famous example when the system in a background magnetic field. In such a case the usual set of plane-wave states would no longer be suitable as a starting point of perturbation. However the standard quantum field theory on textbooks roots in choosing the in and out states are plane-wave which is natural in high energy collisions. Therefore, in a rapidly and globally rotating system, a difficult but straightforward alternative is to reformulate the perturbation computation in principle.  In this work we will complete the first step, which includes solving the Proca equation in present of a background rotation and complete its canonical quantization. It will be shown that because of the symmetry the eigen states are actually the same as the ones of Maxwell equations in cylindrical coordinate\cite{Dai:2012bc,Tobar:2018arx,Chernodub:2018era}. The propagator as well as the near-central approximation will be obtained by considering the vorticity areas are so small in the relativistic QGP. 
We will discuss the subtle zero-mass and gauge-symmetry topics in such a curved space-time system in our further works.

In this work the Latin and Greek index represent components of vectors/tensors in the local rest/inertial frame and curved frame respectively, such as $X^a$ and $X^\mu$.
The $X^a$ respect the usual Lorentz transformation, while the $X^\mu$s obey the coordinate transformation in its curved space-time.
For any vector these two kinds of components are connected by tetrads $e^a_{\ \mu}$ as
\begin{eqnarray}
X^a=e^a_{\ \mu}X^\mu
\end{eqnarray}

\section{Rotating Proca field}
We consider the rotating system in a local rest/inertial frame, up to a local Lorentz transformation tetrads $e_{a}^{\ \mu}$ and $e^{a}_{\ \mu}$
which bridge the vectors in curved and flat space-time can be chosen as
\begin{eqnarray}
e_{\ \mu}^{a}=\left(
\begin{array}{cccc}
 1 & 0 & 0 & 0 \\
 v_1 & 1 & 0 & 0 \\
 v_2 & 0 & 1 & 0 \\
 v_3 & 0 & 0 & 1 \\
\end{array}
\right),
e^{\ \mu}_{a}=\left(
\begin{array}{cccc}
 1 & -v_1 & -v_2 & -v_3 \\
 0 & 1 & 0 & 0 \\
 0 & 0 & 1 & 0 \\
 0 & 0 & 0 & 1 \\
\end{array}
\right)
\end{eqnarray}

By definition they are connected with the global metrics and should satisfy the following relations which roughly means defining a local Minkowski frame by absorbing 
the space-time curvature into local coordinates. 
\begin{eqnarray}
\label{ee1}
&&\eta_{ab}=g_{\mu\nu}e_{a}^{\ \mu}e_{b}^{\ \nu}\\
\label{ee2}
&&e^a_{\ \mu}e^{\ \mu}_{b}=\delta^a_{b}\\
\label{ee3}
&&e^{\ \mu}_{a}e^a_{\ \nu}=\delta^\mu_{\nu}
\end{eqnarray}
where $\eta_{ab}=\{+, -, -, -\}$ and the rotating metric is
\begin{eqnarray}
g_{\mu\nu}=\left(
\begin{array}{cccc}
 1-v_1^2-v_2^2-v_3^2 & -v_1 & -v_2 & -v_3 \\
 -v_1 & -1 & 0 & 0 \\
 -v_2 & 0 & -1 & 0 \\
 -v_3 & 0 & 0 & -1 \\
\end{array}
\right)
\end{eqnarray}

Obviously there are infinite equivalent choices of tetrads satisfying the relation \ref{ee1},\ref{ee2},\ref{ee3} up to a local Lorentz transformation.
Here we consider the uniformly rotation case which means the linear velocity is $\vec{v}=\vec{\omega}\times\vec{x}$,
where $\vec\omega$, the angular velocity, is a constant. For such a globally rotating system the boundary condition is necessary because of the light-speed limit. As the sharp cut boundary will only result in a discrete set of radial wave function thus import no structural changes to the following formalism, we will not do this discussion explicitly in this work and keep the summation over $k_t$ as integration. We should keep in mind for different boundary condition this will be discretized into different series.
And realistically in the QGP the rotation areas or volumes are so small that the light-speed limit problem is far from important\cite{Jiang:2016woz}. 

We will extract rotation effects by construct the Proca Lagrangian density with the field in the local inertial frame. According to the general relativistic principle it should be the same as the flat one except all the quantities are in the curved space-time. By using the relation \ref{ee1} the vector Lagrangian density could be written as
\begin{eqnarray}
\mathcal{L}_{v}=-\frac{1}{4}F^{\mu\nu}F_{\mu\nu}+\frac{1}{2}m^2 A^2=-\frac{1}{4}F^{ab}F_{ab}+\frac{1}{2}m^2 A^2
\end{eqnarray}
where the covariant derivative in the flat space-time is
\begin{eqnarray}
\label{pd}
D_a A_b=e_a^{\ \mu}D_\mu A_b=e_a^{\ \mu}(\partial_\mu A_b+\Gamma_{\mu b c}A^c)
\end{eqnarray}
and the tensor of field strength is given by $F_{ab}=D_a A_b-D_b A_a$.
The reason we need Eq.\ref{pd} is that we are trying to study the system in an inertial frame which means what we observe or measure are the quantities constructed by the field in the flat space-time. The operation $D_a$ can be defined via the $D_\mu$ which operates on the quantities as functions of global coordinates, i.e. Eq.\ref{pd}. By this space-time shifts we are now sitting in a local-rest frame(up to a local Lorentz transformation) and the rotation effects whose effect is bending or twisting the global space-time is felt as some background interactions appearing in the local Lagrangian density as an additional term, e.g. the explicit polarization effect term for the fermionic case.   Here one should note that the connection $\Gamma_{\mu b c}$ is not the usual Christoffel connection. It has included derivatives of the local
inertial basis along the world line of the curved space-time. It is expressed as
\begin{eqnarray}
&&\Gamma_{\mu a b}=\eta_{ac}[e^c_{\ \nu} e^{\ \lambda}_b G^\nu_{\ \mu\lambda}-e^{\ \lambda}_b\partial_\mu e^c_{\ \lambda}]\nonumber\\
&&=\eta_{ac}(\delta^c_{\ \nu}+\xi^c_{\ \nu})(\delta^{\ \lambda}_b+\xi^{\ \lambda}_b)\frac{1}{2}g^{\nu\alpha}(\partial_\mu h_{\alpha\lambda}+\partial_\lambda h_{\alpha\mu}-\partial_\alpha h_{\mu\lambda})\nonumber\\
&&\ \ \ \ -\eta_{ac}(\delta^{\ \lambda}_b +\xi^{\ \lambda}_b)\partial_\mu(\delta^c_{\ \lambda}+\xi^c_{\ \lambda})\nonumber\\
&&=\eta_{ac}(\delta^c_{\ \nu})(\delta^{\ \lambda}_b)\frac{1}{2}\eta^{\nu\alpha}(\partial_\mu h_{\alpha\lambda}+\partial_\lambda h_{\alpha\mu}-\partial_\alpha h_{\mu\lambda})\nonumber\\
&&\ \ \ \ -\eta_{ac}(\delta^{\ \lambda}_b)\partial_\mu(\xi^c_{\ \lambda})\nonumber\\
&&=\frac{1}{2}(\partial_\mu h_{a b}+\partial_b h_{a \mu}-\partial_a h_{\mu b})-\partial_\mu \xi_{a b}
\end{eqnarray}
where $G^\nu_{\ \mu\lambda}$ is the usual Christoffel connection in the curved space-time. Its nonzero elements are
\begin{eqnarray}
&&\Gamma_{0 i j}=\frac{1}{2}(-\partial_j v_i+\partial_i v_j)=\epsilon_{ijm}\omega_m\nonumber\\
&&\Gamma_{i 0 j}=-\frac{1}{2}(\partial_j v_i+\partial_i v_j)\nonumber\\
&&\Gamma_{i j 0}=\frac{1}{2}(\partial_j v_i+\partial_i v_j)
\end{eqnarray}
Obviously the last two equations are actually zero in the uniform rotation case.

Substituting the covariant derivation into the Lagrangian density, the straightforward calculation shows in the local flat frame the field strength tensor is modified by the rotation as
\begin{eqnarray}
&&F_{0i}=-F^{0i}=(\partial_0 A_i-\partial_i A_0)-v_j\partial_j A_i-\epsilon_{ijk}\omega_k A_j\nonumber\\
&&F_{ij}=F^{ij}=\partial_i A_j-\partial_j A_i
\end{eqnarray}
With our choice of the tetrads only the "electric" parts is changed while the "magnetic" parts are invariant. Obviously this conclusion is tetrads-dependent because the "electric" and "magnetic" part are transferable under a local Lorentz transformation.
Hence it is easy to compute the rotation-modified Lagrangian density and split it into the free and rotation-polarization parts as
\begin{eqnarray}
&&\mathcal{L}_{v}(\vec\omega)=\mathcal{L}_{v}(\vec\omega=0)+\delta\mathcal{L}_{v}(\vec\omega)\\
&&=\mathcal{L}_{v}(\vec\omega=0)-v_j f_{j i}f_{0i}+\frac{1}{2}(v_j\partial_j A_i+\epsilon_{i j m}\omega_m A_j)^2\nonumber
\end{eqnarray}
where $\mathcal{L}_{v}(\vec\omega=0)=-\frac{1}{2}(\sum_i-f_{0i}^2+\sum_{i<j}f_{ij}^2)$ and $f_{a b}=\partial_a A_b-\partial_b A_a$.
Here we have lowered all of the indices. It is easy to discover that the $\mathcal{O}(\omega)$ term is polarization form $\vec{v}\cdot(\vec{E}\times\vec{B})=\vec{\omega}\cdot\vec{J}$,
where angular momentum is $\vec{J}=\vec{r}\times \vec{P}$ and $\vec{P}=\vec{E}\times\vec{B}$. However different from the fermionic case the $\mathcal{O}(\omega^2)$ terms are non-negligible because of the eigen equations of 
the vector field are second order, i.e. $\partial_0^2$. The $\mathcal{O}(\omega^2)$ terms are actually corresponding to the $\mathcal{O}(\omega)$ corrections to the eigen energies. This is shown as follows.

The corresponding equations of motion are obtained as 
\begin{eqnarray}
&&\partial_i f_{i 0}-m^2 A_0=\Delta A_0-m^2 A_0=0\\
&&\partial^2_0 A_i-\Delta A_i -2v_j \partial_0\partial_j A_i+(\partial_j v_i-\partial_i v_j)\partial_0 A_j\nonumber\\
\label{roteq}
&&+v_j\partial_j(v_n\partial_n A_i+2\epsilon_{i n m}\omega_m A_n)-(\omega^2 A_i-\omega_n \omega_i A_n)\nonumber\\
&&+m^2 A_i=0
\end{eqnarray}
There are three polarization components for the massive vector field. We could first adopt the transverse constraint $\nabla\cdot\vec{A}=0$ to extract the two transverse ones. Thus the $\partial_i f_{i0}$ is reduced to $\Delta A_0$. Considering the operator $\Delta=\nabla^2$ is semi-positive, the first equation gives $A_0=0$. By choosing the angular velocity direction as the z axis, i.e. $\vec\omega=\omega e_z$ in the cylindrical coordinate the second one degenerates to the ordinary form in the $\omega=0$ case $\partial^2_0 A_i-\Delta A_i=0$. The solutions are the well-known cylindrical waves which could be found in any electrodynamics textbook. Adopting the same symbols as a recent literature \cite{Chernodub:2018era} in the cylindrical system the solutions are
\begin{eqnarray}
&&A_{TE}=\left(
\begin{array}{c}
A_0\\
A_{\rho}^{TE}  \\
A_{\phi}^{TE}  \\
A_{z}^{TE}  \\
\end{array}
\right) =\left(
\begin{array}{c}
0\\
 \frac{n}{k_t\rho}J_{n}(k_t\rho) \\
 \frac{i}{k_t}\partial_\rho J_n(k_t\rho) \\
 0 \\
\end{array}
\right) e^{i (n\phi+ k_z z- E_{n k_t k_z} t)}\nonumber\\
\\
&&A_{TM}=\left(
\begin{array}{c}
A_0\\
A_{\rho}^{TM}  \\
A_{\phi}^{TM}  \\
A_{z}^{TM}  \\
\end{array}
\right) =\left(
\begin{array}{c}
0\\
 \frac{k_z}{E_k k_t}\partial_\rho J_n(k_t\rho) \\
 i\frac{n k_z}{E_k k_t\rho}J_{n}(k_t\rho) \\
 -i\frac{k_t}{E_k}J_n(k_t\rho) \\
\end{array}
\right)e^{i (n\phi+ k_z z- E_{n k_t k_z}t)}\nonumber
\end{eqnarray}
where $E_k=\sqrt{k_t^2+k_z^2}$. In the static case the eigen energies are $E_{n k_t k_z}=E_k$. For the longitudinal part we choose $A_\rho=A_\phi=0$ and the equation gives the solution for the third polarization direction 
\begin{eqnarray}
A_L=\left(
\begin{array}{c}
 \frac{k_z}{E_{kt}}\\
 0 \\
 0 \\
 -\frac{E_k}{E_{kt}} \\
\end{array}
\right)J_n(k_t\rho) e^{i n\phi}e^{i k_z z}e^{-i E_{n k_t k_z} t}
\end{eqnarray}
where $E_{kt}=\sqrt{k_t^2+m^2}$.

As the uniform rotation preserves the cylindrical symmetry we expect these solutions are eigen states of the rotation case as well with a modified energy. By substituting them into Eq.\ref{roteq} we find the equations are satisfied if the energy is solved from
\begin{eqnarray}
(-E_{n k_t k_z}^2+E_k^2-2E_{n k_t k_z}\omega n-\omega^2 n^2)\vec{A}=0
\end{eqnarray}
This equation gives the new eigen energies are $E_{n k_t k_z}=E_k-n\omega$ which as our expectation have obtained $n\omega$ polarization corrections for the spin-1 field on top of the rotation background. 

The nontrivial difference between these solutions and the usual plane-wave case is the following properties are satisfied
\begin{eqnarray}
\label{cur1}
&&\nabla\times\vec{A}_{TE}=E_k \vec{A}_{TM}\\
\label{cur2}
&&\nabla\times\vec{A}_{TM}=E_k \vec{A}_{TE}\\
\end{eqnarray}
As comparison we list the similar relations for the plane-wave $\vec{A}^{pw}_{1,2}$ as follows
\begin{eqnarray}
&&\nabla\times\vec{A}_{1}^{pw}=i|\vec k||\vec{A}^{pw}_1|/|\vec{A}^{pw}_2| \vec{A}^{pw}_1\\
&&\nabla\times\vec{A}_{2}^{pw}=-i|\vec k||\vec{A}^{pw}_2|/|\vec{A}^{pw}_1| \vec{A}^{pw}_2
\end{eqnarray}
The same signs in Eq.\ref{cur1} and \ref{cur2} allow us to construct two new modes, i.e. $\vec{A}_{\pm}=\vec{A}_{TM}\pm\vec{A}_{TE}$,
whose $\vec{E}\cdot\vec{B}\neq 0$. Obviously they satisfy
\begin{eqnarray}
&&\nabla\times\vec{A}_{+}=E_k \vec{A}_{+}\\
&&\nabla\times\vec{A}_{-}=-E_k \vec{A}_{-}\\
\end{eqnarray}
To see this clearly we list the $\vec{E}$ and $\vec{B}$ as follows
\begin{eqnarray}
&&\vec{E}_+=\partial_0 A^+_i=-i E_{n k_t k_z}A^+_i\\
&&\vec{B}_+=-\epsilon_{i j k} \partial_j A^+_k=-E_k A^+_i
\end{eqnarray}
and
\begin{eqnarray}
&&\vec{E}_-=\partial_0 A^-_i=-i E_{n k_t k_z}A^-_i\\
&&\vec{B}_-=-\epsilon_{i j k} \partial_j A^-_k=E_k A^-_i
\end{eqnarray}
Of course this will not cause problems because of $\vec{E}_+\cdot\vec{B}_+ + \vec{E}_-\cdot\vec{B}_- = 0$.

\section{Propagator}
In this section we will complete the canonical quantization for the spin-1 vector field. First we check the orthogonality of the three eigen states. Obviously the $A_{TM}$ has mixed part of $A_L$. In order to construct the propagator of vector field, we first orthogonalize it as 
\begin{eqnarray}
&&B_L=A_L-\alpha A_{TM}\\
&&B_L\cdot A_{TM}=0
\end{eqnarray}
The orthogonalization constraint gives $\alpha=-i k_t\sqrt{k^2+m^2}/(k \sqrt{k^2_t+m^2})$. In the following we will rename $B_L$ as $A_L$.
And all the three eigen states are normalized as
\begin{eqnarray}
&&\int\rho d\rho d\phi dz A^\mu_{TE,TM, L} A_{\mu, TE, TM, L}\nonumber\\
&&=-(2\pi)^2\delta_{mn}\delta(k_z-p_z)\frac{1}{k_t}\delta(k_t-p_t)
\end{eqnarray}
The canonical quantization gives the vector field as
\begin{eqnarray}
A_\mu=\sum_{n, \lambda}\int \frac{d k_z d k_t}{2\pi \sqrt{2E_k}} e^{-i E_n t}e^{i k_z z}e^{i n \phi} a_{n k_t k_z} A_{\mu, \lambda}+h.c.
\end{eqnarray}
where $\lambda\in\{L, TM, TE\}$ is the polarization index. Starting from the quantization assumption $[A_\mu(x), \Pi_\nu(y)]=g_{\mu\nu}\delta^{4}(x-y)$, the commutation relation of $a$s and $a^\dagger$ are derived as
\begin{eqnarray}
[a_{n k_t k_z}, a^\dagger_{m p_t p_z}]=\delta_{m n}\delta(k_z-p_z)k_t\delta(k_t-p_t)
\end{eqnarray}
Therefore the propagator from $x=(t, \vec{x})=(t, \rho, \theta, \phi, z)$ to $y=(s, \vec{y})=(s, r, \theta, \zeta)$ is defined as
\begin{eqnarray}
&&D_{\mu\nu}(x, y)=\langle 0|T A_\mu(x)A_\nu(y)|0\rangle\\\nonumber
&&=\theta(t-s)\langle 0|A_\mu A_\nu|0\rangle+\theta(s-t)\langle 0|A_\nu A_\mu|0\rangle
\end{eqnarray}
where the $T$ is the time-ordering operator. Substituting the field operators into it and changing the summation over $n$ into $-n$ in one of the two terms, the propagator is reduced as
\begin{eqnarray}
&&D_{\mu\nu}(x, y)=\frac{i}{(2\pi)^3}\sum_{n, \lambda}\int d k_0 d k_z k_t d k_t e^{i k_z (z-\zeta)}e^{i n (\phi-\theta)}\nonumber\\
&&\times e^{-i(k_0-n\omega)(t-s)} \frac{ A_{\mu,\lambda}(k_t, n, k_z;\rho)A^*_{\nu,\lambda}(k_t, n, k_z;r)}{k_0^2-E_k^2+i\eta}\nonumber\\
&&=\frac{i}{(2\pi)^3}\sum_{n, \lambda}\int d k_0 d k_z k_t d k_t e^{i k_z (z-\zeta)}e^{i n (\phi-\theta)}\nonumber\\
&&\times e^{-i k_0(t-s)} \frac{ A_{\mu,\lambda}(k_t, n, k_z;\rho)A^*_{\nu,\lambda}(k_t, n, k_z;r)}{(k_0-n\omega)^2-E_k^2+i\eta}
\end{eqnarray}
which is similar to the result in flat space-time. At the second equation we have shifted the integration over $k_0$ to $k_0-n\omega$ to make the poles locate at $E_k-n\omega$ which corresponding to the eigen energies of the vector field. Once again we find the rotation effects serve as an additional chemical potential rather than a straightforward energy shift as $k_0^2-E_{n k_t k_z}^2$. This is the same as the fermionic case.  And the straightforward computation gives the Lorentz structure part as
\begin{eqnarray}
&&D^{n}_{\mu\nu}(k_t, k_z;\rho, r)=\sum_\lambda A_{\mu,\lambda}(k_t, n, k_z;\rho)A^*_{\nu,\lambda}(k_t, n, k_z;r)\nonumber\\
&&=\frac{1}{4}\left(
\begin{array}{cccc}
 0 & 0 & 0 & 0 \\
 0 & M_n^{++} & -i M_n^{+-} & 0 \\
 0 & i M_n^{-+} & M_n^{--} & 0 \\
 0 & 0 & 0 & 0 \\
\end{array}
\right)\\
&&+\frac{E_{kt}^2}{4 m^2}\left(
\begin{array}{cccc}
 0 & 0 & 0 & 0 \\
 0 &  M_n^{--} & -i M_n^{-+} & -2i\frac{k_t k_z}{E_{kt}^2}N_n^{1-} \\
 0 & i M_n^{+-} &  M_n^{++} & 2\frac{k_t k_z}{E_{kt}^2}N_n^{1+} \\
 0 & 2i\frac{k_t k_z}{E_{kt}^2}N_n^{2-} & 2\frac{k_t k_z}{E_{kt}^2}N_n^{2+} & 4\frac{E_z^2}{E_{kt}^2}\Pi_n\\
\end{array}
\right)\nonumber\\
&&+\frac{k_t E_{kt}}{2m^2}
\left(
\begin{array}{cccc}
 \frac{2k^2}{k_t E_k}\Pi_n & -i N_n^{2-} & -N_n^{2+} & -2\frac{k_z}{k_t}\Pi_n \\
 i N_n^{1-} & 0 & 0 & 0 \\
 - N_n^{1+} & 0 & 0 & 0 \\
 -2\frac{k_z }{k_t}\Pi_n & 0 & 0 & 0 \\
\end{array}
\right)\nonumber
\end{eqnarray}
In order to make the expression more compact we have used symbols
\begin{eqnarray}
M_n^{++}=Z^+_n(\rho,\phi)Z^+_n(r, \theta)\nonumber\\
M_n^{+-}=Z^+_n(\rho,\phi)Z^-_n(r, \theta)\nonumber\\
M_n^{-+}=Z^-_n(\rho,\phi)Z^+_n(r, \theta)\nonumber\\
M_n^{--}=Z^-_n(\rho,\phi)Z^-_n(r, \theta),
\end{eqnarray}
\begin{eqnarray}
N_n^{1+}=Z^+_n(\rho,\phi)J_n(r) \nonumber\\
N_n^{1-}=Z^-_n(\rho,\phi)J_n(r)\nonumber\\
N_n^{2+}=J_n(\rho)Z^+_n(r,\theta)\nonumber\\
N_n^{2-}=J_n(\rho)Z^-_n(r, \theta)
\end{eqnarray}
and
\begin{eqnarray}
\Pi_n=J_n(\rho)J_n(r)
\end{eqnarray}
where $Z^+_n(\rho, \phi)=J_{n-1}(\rho)e^{-i\phi}+J_{n+1}(\rho)e^{i\phi}$ and 
$Z^-_n(\rho, \phi)=J_{n-1}(\rho)e^{-i\phi}-J_{n+1}(\rho)e^{i\phi}$ and write the $J_n(k_t \rho)$ as $J_n(\rho)$ for simplicity. Beware here the propagator is in the Cartesian coordinate, e.g. the $D^n_{01}=D^n_{tx}$ and so on. Obviously the propagator is not invariant under translation on the $X-Y$ plane. If we set the rotation speed $\omega=0$ it will be reduced to the usual one $D_{\mu\nu}(x-y)$ in flat space-time. This can be checked by a straightforward computation. During the checking process the following relations are helpful
\begin{eqnarray}
&&\sum_n e^{in(\phi-\phi')}J_n(k_t\rho)J_n(k_t\rho')=J_0(k_t|\vec{\rho}-\vec{\rho}'|)
\end{eqnarray}

\begin{eqnarray}
&&\int_0^{2\pi}d\xi e^{ikr cos\xi}cos(\xi+\alpha)=2\pi i cos\alpha J_1(k r)\nonumber\\
&&\int_0^{2\pi}d\xi e^{ikr cos\xi}sin(\xi+\alpha)=2\pi i sin\alpha J_1(k r)
\end{eqnarray}

\begin{eqnarray}
&&\int_0^{2\pi}d\xi e^{ikr cos\xi}cos(\xi+\alpha)cos(\xi+\alpha)\nonumber\\
&&=\pi J_0(k r)-\pi cos2\alpha J_2(k r)\nonumber\\
&&\int_0^{2\pi}d\xi e^{ikr cos\xi}sin(\xi+\alpha)sin(\xi+\alpha)\nonumber\\
&&=\pi J_0(k r)+\pi cos2\alpha J_2(k r)\nonumber\\
&&\int_0^{2\pi}d\xi e^{ikr cos\xi}cos(\xi+\alpha)sin(\xi+\alpha)\nonumber\\
&&=-\pi sin2\alpha J_2(k r)
\end{eqnarray}

\section{Near central expansion}
For a globally rotating system, the size is limited, especially for the QGP which is believed rotating very rapidly fireball product in relativistic heavy ion collisions. AMPT's simulation indicates the vorticity profile is more like a set of vorticity spots rather than a rotating lemon. Therefore we consider the propagator in the near central case and set one of the space-time point as the origin, i.e. $r=0$ and $\rho\rightarrow 0$. This will reduce it into three non-zero parts corresponding to $n=0, \pm 1$ because of only $J_0(0)\neq 0$. The explicit results are
\begin{widetext}
\begin{eqnarray}
&&D^{n=-1}_{\mu\nu}(k_t, k_z, \rho)
=\frac{1}{4}\left(
\begin{array}{cccc}
 0 & -2i\frac{k_t E_k}{m^2} J_1 & 2\frac{k_t E_k}{m^2} J_1 & 0 \\
 0 & 2J_0 e^{i\phi}-\frac{k_t^2}{m^2}Z^-_{-1}(\rho,\phi) & 2 i J_0 e^{i\phi}-i \frac{k_t^2}{m^2}Z^-_{-1}(\rho,\phi) & 0 \\
 0 & -2 i J_0 e^{i\phi}-i\frac{k_t^2}{m^2}Z^+_{-1}(\rho,\phi) & 2J_0 e^{i\phi}+\frac{k_t^2}{m^2}Z^+_{-1}(\rho,\phi) & 0 \\
 0 & 2 i \frac{k_t k_z}{m^2}J_1 & -2 \frac{k_t k_z}{m^2}J_1 & 0 \\
\end{array}
\right)\nonumber
\end{eqnarray}

\begin{eqnarray}
&&D^{n=0}_{\mu\nu}(k_t, k_z, \rho)
=\left(
\begin{array}{cccc}
 \frac{k^2}{m^2}J_0 & 0 & 0 & -\frac{E_k k_z}{m^2} J_0 \\
 i\frac{k_t E_k}{2m^2}Z^-_{0}(\rho,\phi) & 0 & 0 & -i\frac{k_t E_k}{2m^2}Z^-_{0}(\rho,\phi) \\
 -\frac{k_t E_k}{2m^2}Z^+_{0}(\rho,\phi) & 0 & 0 & \frac{k_t E_k}{2m^2}Z^+_{0}(\rho,\phi) \\
 -\frac{E_k k_z}{m^2}J_0 & 0  & 0 & \frac{E^2_z}{m^2}J_0 \\
\end{array}
\right)
\end{eqnarray}

\begin{eqnarray}
&&D^{n=1}_{\mu\nu}(k_t, k_z, \rho)
=\frac{1}{4}\left(
\begin{array}{cccc}
 0 & -2i\frac{k_t E_k}{m^2} J_1 & -2\frac{k_t E_k}{m^2} J_1 & 0 \\
 0 & 2J_0 e^{-i\phi}+\frac{k_t^2}{m^2}Z^-_{1}(\rho,\phi) & -2 i J_0 e^{-i\phi}-i \frac{k_t^2}{m^2}Z^-_{1}(\rho,\phi) & 0 \\
 0 & 2 i J_0 e^{-i\phi}+i\frac{k_t^2}{m^2}Z^+_{1}(\rho,\phi) & 2J_0 e^{-i\phi}+\frac{k_t^2}{m^2}Z^+_{1}(\rho,\phi) & 0 \\
 0 & 2 i \frac{k_t k_z}{m^2}J_1 & 2 \frac{k_t k_z}{m^2}J_1 & 0 \\
\end{array}
\right)\nonumber
\end{eqnarray}
\end{widetext}

\section{Summary and Outlook}
We studied the massive vector field with the Proca equation in the present of rotation background by introducing the curved metrics in the equation. Different from the fermionic case besides a explicit polarization term $\vec{\omega}\cdot\vec J$ there are $\mathcal{O}(\omega^2)$ terms which are non-negligible as well. As our expectation because of the cylindrical symmetry in the rotating system the eigen states are the same as the free Proca equation in this coordinates. And the polarization effects are also obtained as the energy shift in the eigen energies as $\Delta E=n\omega$. With canonical quantization the propagator is also computed. Its poles indicate that the rotation serves as an additional chemical potential rather than a trivial energy modification. This is the same as that in the fermionic case. It is also shown that the translation symmetry has been broken in the cylindrical coordinate and thus the propagator depends on two space-time points explicitly. Such a behavior will obviously make the further loop computation much more complicated. As a potential simplified case we studied the near-central result briefly because of the size of rotating area/volume is usual limited. We have checked in zero-rotation limit it can be reduced to the usual form explicitly by using the summation relations of Bessel functions. 

This work is a preliminary study on the complete framework of solving the quantum field problems defined in a curved coordinate. In such a case the translation symmetry will be broken and so will the usual energy-momentum conservation at interaction vertices because of the momentum are no longer good quantum numbers when the space is not flat. In this work we have shown that although the computation is much more complicated the standard process is straightforward for the interaction-free case. But it could be expected that the interaction terms will induce more fundamental problems especially for the p-wave cases, e.g. $A^2\partial A$ which will suffer not only the no-flat eigen states modification but also additional terms from the covariant derivation. Obviously this will appear in the non-abelian gauge field model such as QCD. Therefore before face the interaction problem we should also do more studies on the gauge symmetry in such coordinates. We will continue to discuss these topics in our future works.     

{\bf Acknowledgments.}
The work of this research is supported by the National Natural Science Foundation of China, Grant Nos. 11875002(YJ) and the Zhuobai Program of Beihang University.

\end{document}